\documentclass[12pt,preprint]{aastex}
\def\({\left(}
\def\){\right)}
\def\[{\left[}
\def\]{\right]}
\def\msun{{\rm ~M}_{\odot}}

\usepackage{graphicx}
\usepackage{natbib}
\usepackage[usenames,dvips]{color}

\begin{document}
\title{A Hybrid Two Component Accretion Flow Surrounding Supermassive
Black Holes in AGN}
\author{B. F. Liu\altaffilmark{1},
Ronald E. Taam\altaffilmark{2,3}, Erlin Qiao\altaffilmark{1}, and W. Yuan\altaffilmark{1}}
\altaffiltext{1}{National Astronomical Observatories, Chinese Academy of Sciences, 
20A Datun Road,  Chaoyang District, Beijing 100012, China; bfliu@nao.cas.cn}
\altaffiltext{2}{Academia Sinica Institute of Astronomy and Astrophysics-TIARA, P.O. Box 23-141, 
Taipei, 10617 Taiwan; taam@asiaa.sinica.edu.tw}
\altaffiltext{3}{Department of Physics and Astronomy, Northwestern University,  2131 Tech Drive, 
Evanston, IL 60208, USA}

\begin{abstract}

It is commonly believed that the optical/UV and X-ray emissions in luminous AGN are produced in an accretion 
disk and an embedded hot corona respectively. The inverse Compton scattering of disk photons by hot electrons in 
the corona can effectively cool the coronal gas if the mass supply is predominantly via a cool disk like 
flow as in BHXRBs.  Thus, the application of such a model to AGNs fails to produce their observed X-ray 
emission. As a consequence, a fraction of disk accretion energy is usually assumed to be transferred to the 
corona.  To avoid this assumption, we propose that  gas in a vertically extended 
distribution is supplied to a supermassive black hole by the gravitational capture of interstellar medium or stellar wind 
material. In this picture, the gas partially condenses to an 
underlying cool disk as it flows toward the black hole, releasing accretion energy as X-ray emission and 
supplying mass for  the disk accretion. 
Detailed numerical calculations reveal that the X-ray luminosity can reach a few tens of percent of the bolometric 
luminosity.  The value of $\alpha_{\rm ox}$ varies from 0.9 to 1.2 for the mass supply rate ranging from 0.03 to 0.1 times the Eddington value. 
The corresponding photon index in the 2-10 keV 
energy band varies from 1.9 to 2.3.  Such a picture provides a natural 
extension of the model for low luminosity AGN where condensation is absent 
at low mass accretion rates  and no optically thick disk exists in the inner region.
\end{abstract}

\section{Introduction}

Accretion of gas onto a supermassive black hole is central to our understanding of active galactic nuclei (AGN).  
Observational data commonly point to the co-existence of hot and cool accretion flows, which have been described 
in terms of either a hot geometrically thick corona lying above and below a cold standard geometrically thin disk or an inner 
advection dominated accretion flow (ADAF) connected to an outer thin disk.  For high luminosity AGNs 
(HLAGNs), the optical/UV and X-ray emissions are widely believed to originate from a thin disk and an 
overlying hot corona respectively.  On the other hand, for low luminosity AGN (LLAGN), the radiation is commonly 
speculated to arise from an ADAF.

Theoretically, the nature of the accretion process is thought to be similar for both stellar mass black holes 
in X-ray binaries and supermassive black holes in the nuclei of galaxies.  The accretion model successfully 
developed for the black hole X-ray binaries (BHXRBs) is often applied to AGN. However, the model involving the 
coexistence of a disk and corona as applied to luminous, radio quiet AGNs is inadequate. 
In particular, the inverse 
Compton scattering of photons from the geometrically thin disk by hot electrons in the corona quickly cools the 
coronal gas, assuming the mass supply to the accretion is dominated by a disk-like flow as in BHXRBs. In this 
case, the corona is too weak to emit sufficiently high X-ray radiation (Haardt \& Maraschi 1991;Nakamura \& Osaki 1993; Kusunose \& Mineshige 1994;  \.Zycki et al. 1995;  Witt et al. 1997; Liu et al. 2012; Meyer et al. 2012).  
This is in contrast to the observational spectra showing that the coronal fraction in luminous AGNs can often be 
as high as a few tens of percent (e.g., Mushotzky et al. 1993; Elvis et al. 1994; Yuan et al. 1998), 
significantly higher than BHXRBs in the soft state. To alleviate this problem, a fraction of disk accretion 
energy is often assumed to be transferred to the corona (e.g. Haardt \& Maraschi 1991, Nakamura \& Osaki 1993; 
Svensson \& Zdziarski 1994;  \.Zycki et al. 1995;  Witt et al. 1997; Dove et al. 1997; Kawaguchi et al. 2001; Czerny et al. 2003;  Liu et al. 2012). 

Recently, progress has 
been achieved in establishing an accretion disk corona via MHD simulations (e.g., Miller \& Stone 2000; 
Hirose et al. 2006; Guan \& Gammie 2011; Simon et al. 2011, 2012; Bai \& Stone 2013; Fromang et al. 2013). 
In addition, Uzdensky (2013) has developed an analytical model in which the local turbulent energy 
dissipation rate is nearly uniform in the vertical direction.  Such a profile can result in the release of a 
few tens of percent of the accretion power in the corona by the Parker instability for a gas-pressure-dominated 
disk.  We note that there is no necessity for such coronal heating in BHXRBs in the soft state and, hence, 
the vertical dissipation rate profile may be significantly different in the radiation-pressure-dominated 
disk (Blaes et al. 2011), which is the case for luminous AGNs.  A strong corona is supported in complementary 
studies based on the disk evaporation driven by  magnetic reconnection heating via the Parker instability 
(Liu et al.  2002a, 2003).  However, the radiative spectra of such disks in the X-ray band do not vary 
significantly with the Eddington ratio. If a high ion temperature is assumed, this spectral issue can be resolved 
(see Cao 2007), however,  the hot gas shall be quickly accreted to the black hole without a steady hot gas supply.

We suggest that the physical properties of the gas fuel can play an important role in clarifying the  
difference in the nature of accretion between AGNs and BHXRBs. In cases where mass transfer from a companion 
star takes place via Roche lobe overflow (RLO) in BHXRBs, the gas is cool and is mostly constrained to 
lie in the orbital plane.  In contrast, the accreting gas in AGN likely reflects the gravitational capture 
of gas from the interstellar medium or from the stellar winds emitted by evolved stars in the central regions 
of a galaxy (Ho 2008). The radiative feedback from a central source is likely 
to result in a gaseous environment characterized by both hot and cold components (e.g., see Wada 2012, 2013).

In such an environment, large differences in the radiation spectrum are possible, if a thin disk exists 
initially. The Bondi accretion flow is hot as a fraction of the gravitational potential energy is converted into internal 
energy due to compressional heating. As the 
gas flows   via an ADAF toward the black hole, the hot gas partially condenses to the underlying disk as a 
consequence of a disk corona interaction. As a certain amount of hot gas can remain in the corona, it can 
enhance the X-ray emission.  On the other hand, the condensation of the hot gas supplies gas for the accretion 
via a cooler thin disk. We show that the hot and cold accretion rates vary with radius and that 
the flows can significantly contribute to the X-ray emission above that predicted based on accretion 
flows without such a hot component.
In \S 2, we outline the basic physics and assumptions of the interaction 
between the disk and corona.  Assuming a hot gas supply to the black hole in AGN, we present the  
properties of the accretion flows and emission in \S 3. The implications of the model are discussed in \S 4.

\section{Physical Description and Assumptions of the Disk and Corona Model}

   Consider an accretion state in which a geometrically thin, cold disk initially surrounds a black hole. The thin disk
can achieve a steady state if there is a continuous mass supply for the disk accretion.  For stellar mass 
black holes, such a state can be realized since the mass is transferred from a companion star via RLO 
to the thin disk as in the case for BHXRBs. On the other hand, for the case of supermassive black holes, the mass 
supply to the disk results from the gravitational capture of stellar wind or interstellar medium material by the central 
black hole. Such gas is not necessarily constrained to lie in the disk plane, and hence can form a hot accretion 
flow surrounding the disk as a consequence of inefficient radiative cooling in a geometrically thick flow.  As such, 
  we study the interaction between a geometrically thin, optically thick,  pre-existing disk and a geometrically thick, optically thin,  hot accretion flow in such a context. 
If the interaction results in mass evaporating from the disk to the corona, the initial disk is evacuated when the 
accretion flow reaches a steady state.  In contrast, if condensation occurs, the thin disk can be maintained in a 
steady state with an accretion rate equal to the condensation rate.  Since the evaporation/condensation feature 
depends on the distance from the black hole, the thin disk is expected to be described as stationary only in a 
limited range in radial scale. In the following, we investigate such accretion flows to a non-rotating black hole.  

\begin{figure*}
\includegraphics[width=150mm,height=90mm,angle=0.0]{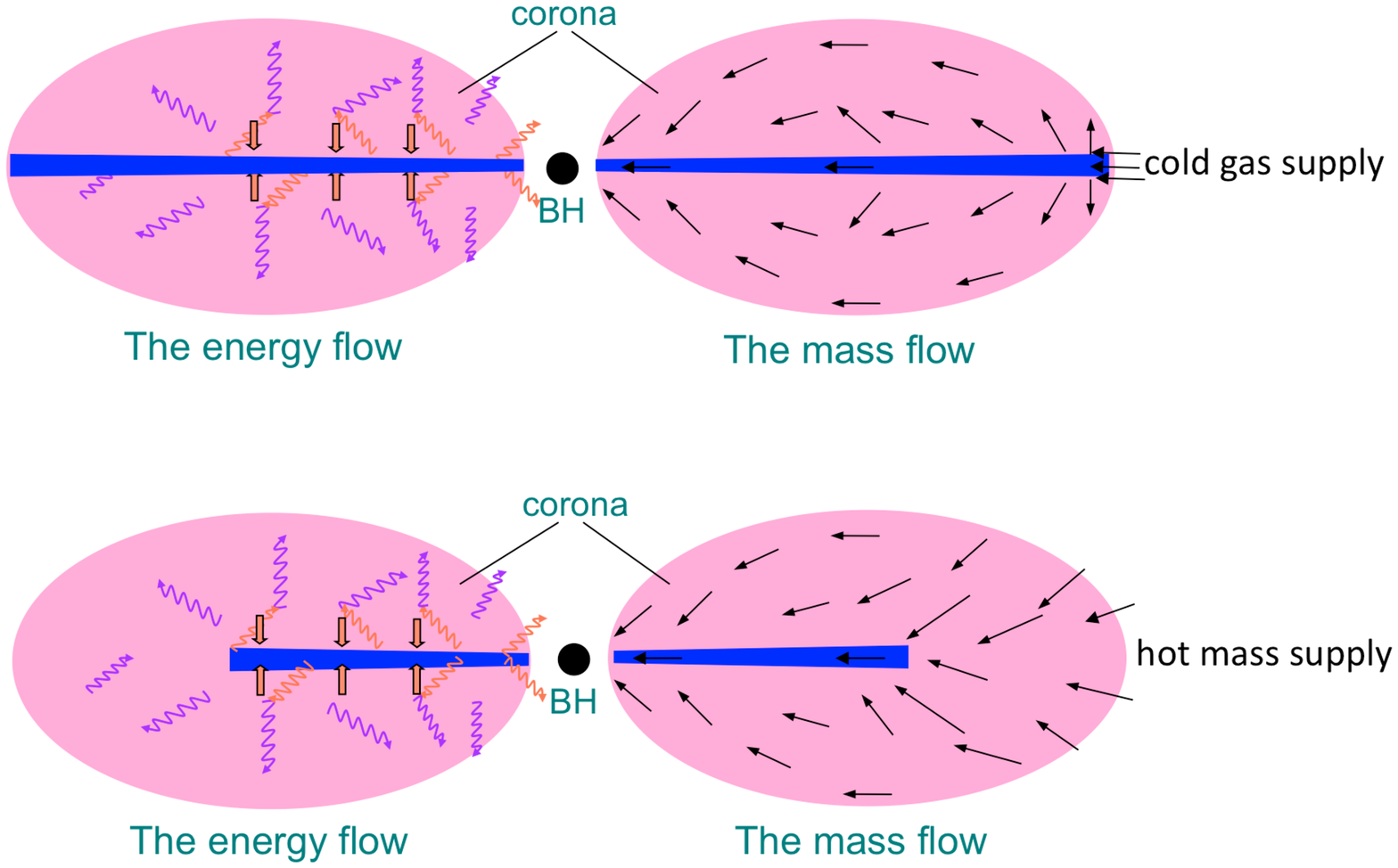}
\caption{\label{f:disk-corona}} 
A schematic description of the mass 
and energy flowing in the disk and corona with cold Roche lobe overflow (upper) and hot, 
isotropically distributed gas supply (lower). 
\end{figure*}

A schematic description of the mass and energy flow in a disk and corona with a cold or hot gas supply is 
displayed in the upper and lower panel of Figure \ref{f:disk-corona} respectively. The accretion is assumed to 
initially take place via a standard optically thick and geometrically thin disk (Shakura \& Sunyaev 1973) and a hot, 
geometrically thick,  optically thin accreting corona. Both the disk and corona are individually powered by the release of gravitational 
energy associated with the accretion of matter.  The interaction between the corona and the underlying disk results from  
the vertical conduction of heat by electrons, the inverse Compton scattering (ICS) of photons from the disk and corona by 
coronal electrons, mass exchange and, hence, enthalpy flow.   As a simplified model, we consider a limiting case 
where only an isotropically distributed gas is supplied to accretion, as shown in the lower panel of Figure 
\ref{f:disk-corona}. The physical processes taken into account in this case  are basically the same as for mass supply via RLO. 
However, with an isotropically distributed mass supply to the corona, the description of the accretion flow in a 
stationary state is expected to differ from the stellar mass black hole context in BHXRBs, in particular at high mass 
accretion rates.

Two methods have been used to investigate such accretion flows.  The study of the vertical structure from the 
chromosphere to a typical corona was pioneered by Meyer \& Meyer-Hofmeister (1994) and investigated for black holes 
(Liu et al. 1999; Meyer et al. 2000a,b; Liu et al. 2002b; Liu \& Taam 2009; R\`o\.za\`nska \& Czerny 2000a,b; Mayer 
\& Pringle 2007; Bradley \& Frank 2009; Taam et al. 2012; Liu et al. 2013). Stratification models (Liu et al. 2006; 
2007; Meyer et al. 2007; see also Zhang et al. 2000) have been proposed in a simplified approximation. These studies 
reached the similar conclusion that the mass exchange between the disk and corona is essential.   In this study, 
we adopt a model where the accretion flows are vertically stratified into 3 layers, that is, a thin disk, a transition 
layer and a typical corona layer. 
The corona is taken as a two-temperature flow, which is similar to an ADAF except that there is additional cooling by conduction and external Compton scatterings. The pressure ($p$), electron density ($n_e$), viscous heating rate ($q^+$), and sound speed ($c_s$) in the corona  can be    determined by (Narayan \& Yi 1995b)
\begin{eqnarray}\label{para}
\begin{array}{l}
p=1.71\times10^{16}\alpha^{-1}c_{1}^{-1}c_{3}^{1/2}m^{-1}\dot m_{\rm cor} r^{-5/2} \ \ \ \rm g \ cm^{-1} \ s^{-2},  \\
n_e=2.00\times10^{19}\alpha^{-1}c_1^{-1}c_{3}^{-1/2}m^{-1} \dot m_{\rm cor} r^{-3/2}\ \ \ \rm cm^{-3}, \\
q^{+}=1.84\times 10^{21}\varepsilon^{'}c_{3}^{1/2}m^{-2}\dot
m_{\rm cor} r^{-4}\ \ \rm ergs \ cm^{-3} \ s^{-1} , \\
c_s^2 =4.50\times10^{20}c_{3}r^{-1} \ \ \rm cm^{2} \ s^{-2},
\end{array}
\end{eqnarray}
where $\alpha$ is the viscosity parameter, $m$, $\dot m_{\rm cor}$ and $r$ are the black hole mass  in units of solar mass,  
the coronal accretion rate in units of the Eddington rate ($\dot M_{\rm Edd}=1.39\times 10^{18}m$ g/s), and  the distance 
in units of the Schwarzschild radius, $R_S$, respectively. $c_1$, $c_3$ and $\epsilon '$ depend on the energy fraction of advection   $f$,
\begin{equation}
\begin{array}{l}
{c_1}={(5+2\varepsilon^{'}) \over {3\alpha^2}}g(\alpha,\varepsilon^{'}),\\
{c_3}={2\varepsilon(5+2\varepsilon^{'})\over {9\alpha^2} } g(\alpha,\varepsilon^{'}),\\
{\varepsilon{'}}={\varepsilon\over f}={1\over f} \biggl({{5/3-\gamma}\over {\gamma-1}}\biggr),\\
g(\alpha,\varepsilon^{'})=\biggl[ {1+{18\alpha^2\over (5+2\varepsilon^{'})^{2}}\biggr]^{1/2}-1},\\
\gamma={{32-24\beta-3\beta^2}\over {24-21\beta}}.\\
\end{array}
\end{equation}

  With the pressure and density expressed in Eq.(1), the ion temperature ($T_i$) in the corona is determined by the equation of state.  The electron temperature ($T_e$) is determined by the energy balance between heating  through Coulomb collisions 
and cooling through conduction and radiation by  bremsstrahlung, synchrotron, and  Compton scattering of both the 
corona and disk radiations, that is,
\begin{eqnarray}\label{e:energy}
\Delta F_{\rm c}/H=q_{\rm ie} -q_{\rm rad},
\end{eqnarray}  
where  $q_{\rm ie}$ is the energy transfer from ions to electrons (Stepney 1983),
\begin{eqnarray}
q_{\rm ie} & = & (3.59\times 10^{-32} {\rm g\ cm^5\ s^{-3}\ K^{-1}})
n_e n_i T_i {\left(\frac{k T_e}{m_e c^2}\right)}^{-3/2}.
\end{eqnarray}
$q_{\rm rad}$ is the radiative
cooling rate of the corona including the bremsstrahlung cooling rate, synchrotron
cooling rate and the corresponding self-Compton scattering, which is a function of electron temperature, 
density and energy density of soft photons. H is  the scale height of the corona, given approximately by
$H=(2.5c_{3})^{1/2}rR_{\rm S}$. 
$\Delta F_{\rm c}$ is the net flux flowing out of the corona.  In the case of no flux coming into the corona at its upper boundary, $\Delta F_{\rm c}$ equals to the flux transferred from the corona to the transition layer $F_c^{ADAF}$, that is,
\begin{equation}
\Delta F_{\rm c}=F_c^{ADAF} =k_{0}T_{\rm e}^{5/2}{dT_{\rm e}\over dz}\approx k_{0}T_{\rm e}^{7/2}/H,
\end{equation}
where the temperature in the transition layer is assumed to be much lower than that in the corona.
Therefore, the electron temperature can be determined by the energy equation Eq.(\ref{e:energy}). The radiative coupling is implicitly included in the Compton cooling rate. The coronal quantities, e.g. the electron temperature, the density and 
pressure are all affected by the disk emission.  

Near the base of the corona, the electrons are coupled with the ions. With the decrease of electron temperature from 
this coupling interface to the disk surface, the density undergoes a dramatic increase in the deeper vertical region 
because the pressure, in balance with the vertical component of gravity, does not significantly change in this layer 
compared to the large change in temperature.
As a consequence, bremsstrahlung energy losses become much more important in this layer than in a typical corona.  
How the gas flows vertically in the transition layer strongly depends on the upper corona parameters. If the pressure 
in the corona is high, the density in the transition layer is also high. Hence, the radiative cooling is more efficient 
than the heating associated with the energy flux conducted from the ADAF. As a consequence, a part of the gas condenses 
to the disk, releasing its thermal energy as additional heat. On the other hand, if the pressure is low, only part 
of the conductive flux can be radiated away, and the remaining flux will heat up the cool disk matter, leading to 
matter evaporation from the disk to the corona. As the properties of the corona depend on the accretion rate in the 
corona and photons from the disk, whether evaporation or condensation takes place and the degree to which mass exchange 
is important is dependent on the relative mass supply rate in the two components relative to the total accretion rate. 
In the case, with only mass supply to the corona accretion, the evaporation/condensation rate  can be determined by the 
accretion rate in the corona for a given black hole mass (Meyer et al. 2007; Liu et al. 2007, Taam \& Liu 2008; Qiao 
\& Liu 2013) as    follows.
The energy balance in the
transition layer is set by the incoming conductive flux,
bremsstrahlung radiation flux, and the enthalpy flux carried by the
mass condensation flow,
\begin{equation}\label{energy-layer}
\frac{d}{dz} \left[\dot m_z \frac{\gamma}{\gamma-1}
\frac{1}{\beta}\frac{\Re T}{\mu} + F_c \right] = -n_e n_i
\Lambda(T).
\end{equation}
This yields the evaporation/condensation rate per unit area (Meyer et al. 2007; Liu et al. 2007),

\begin{equation}\label{cnd-general}
\dot m_z= {{\gamma-1} \over \gamma}\beta {{-F_{c}^{ADAF}} \over {\Re
T_{i}/ \mu_{i}}}(1-\sqrt{C}),
\end{equation}
with
\begin{equation}\label{C}
C \equiv\kappa{_0} b \left(\frac{\beta^2 p^2}{\pi k^2}\right)
\left(\frac{T_{\rm {cpl}}}{F_c^{\rm{ADAF}}}\right)^2,
\end{equation}
where $\dot m_z$ denotes the evaporation rate (if positive) or condensation rate (if negative) per unit surface area, 
$T_{\rm cpl}$ the coupling
temperature determined by assuming the heating of the ions is completely transferred to the electrons, 
$\gamma={8-3\beta\over 6-3\beta}$ (Esin 1997).
Other quantities are constant, specifically, $\kappa_0 = 10^{-6}{\rm erg\,s^{-1}cm^{-1}K^{-7/2}}$ related to the 
conductivity (Spitzer 1962), $b=10^{-26.56}\,\rm g\,cm^5s^{-3}deg^{-1/2}$ as related to bremsstrahlung radiation 
(Sutherland \& Dopita 1993), $k$ the Boltzmann constant, and $\mu_i=1.23$ corresponding to the ion weight for an 
assumed chemical abundance of $X=0.75$, $Y=0.25$.

For $C>1$, Eq. (\ref{cnd-general}) yields  $\dot m_z < 0$, indicating that the coronal matter condenses into 
the disk. On the other hand, for $C<1$, mass evaporates from the disk to the corona.  The condition, $C=1$, separates 
the regions of evaporation and condensation. Numerical calculations show that $C$ increases with decreasing distance, 
which is due to more efficient bremsstrahlung radiation in the transition layer resulting from larger densities in 
the inner region (see Eq.\ref{C}). Thus, $C=1$ determines a critical radius $R_{\rm d}$, interior to which part of 
the corona/ADAF matter condenses onto the disk, while exterior to which the disk matter evaporates into the corona.
In the case of mass supply only to the corona, the disk can only exist steadily interior to this critical distance. 
We take this radius as the outer radius of the disk.

  The integrated condensation rate (in units of the Eddington rate) from $R_d$ to an inner radius, $R$, of the disk is 
calculated by combining  eqs. (\ref{cnd-general}) and (\ref{C}), which reads,

\begin{equation}\label{condensation}
\dot m_{\rm cnd}(R)= -2\int_{R}^{R_d} {2\pi R' \over \dot M_{\rm
Edd}} \dot m_z (R')dR'.
\end{equation}

This integrated condensation rate, $\dot m_{\rm cnd}(R)$, is the accretion rate at distance $R$ (where $R<R_d$) of the 
disk.  The mass accretion rate remaining in the inner corona ($R<R_d$) is derived from mass conservation, 
\begin{equation}\label{mdot-corona}
\dot m_{\rm cor}(R)= \dot m -\dot m_{\rm cnd} (R),
\end{equation}  
where $\dot m$ is the mass supply rate to the corona. 

 The calculation of the accretion rates in both the corona and 
the disk and the corona parameters (e.g., temperature, optical depth) is not expressed analytically since they 
depend on the unknown energy density of the soft photons from the disk, which is contributed by irradiation 
from the corona and accretion in the disk, and hence dependent on the condensation rate. Thus, iterative 
calculation is necessary in order to obtain a self-consistent solution. For a given black hole mass, mass 
supply rate to the corona $\dot m$, 
viscosity parameter, magnetic parameter, advection fraction, and albedo, a maximum soft photon field  is assumed to calculate the condensation rate and corona luminosity, from which a new energy density of the soft photons is derived. An iteration is performed until the previously updated density of soft photons is consistent 
with the derived value, corresponding to a self-consistent solution of the disk and corona.  Given this structure, 
the emergent spectrum of the disk and corona system is calculated by a Monte Carlo simulation (similar to the method 
described in Qiao \& Liu 2013).

\section{The accretion flows in AGN}

Given the mass  of a black hole, $M=10^8M_\odot$, viscosity parameter, $\alpha=0.3$, magnetic field parameter (the 
ratio of gas pressure to total pressure), $\beta=0.8$, advection fraction of 0.1
and albedo of 0.15,  we   perform numerical calculations and examine the structure of the hot 
accretion flow. It is shown that for a given accretion rate in the corona there exists a critical distance, $R_d$, 
within which the coronal gas condenses to a cool disk. For distances greater than $R_d$, disk gas will evaporate 
to the corona. With only the capture of isotropically distributed gas supply as assumed here, any existing gas in a thin disk in the outer 
region ($R>R_d$) will be evacuated by the evaporation process and we take $R_d$ as the outer boundary of the disk. The 
tendency for condensation resulting from efficient Compton cooling is mitigated by the thermal energy left in the corona 
by the condensed gas and the continuous hot gas supply preventing collapse of the corona. 
Thus,  the accretion takes place as an ADAF with a 
nearly constant accretion rate in the outer region ($R>R_d$), while in the inner region, a fraction of hot gas 
condenses to the disk as it approaches the black hole.  This leads to a decreasing mass flow rate in the corona 
and a corresponding increase of the flow rate in the cool disk toward the black hole.

Such a picture applies for sufficiently high accretion rates, $\dot m > 0.02$ in the hot flow.  For lower hot 
gas supply rates, $R_d$ is interior to the innermost stable circular orbit (ISCO), implying that the 
coronal gas cannot condense to a disk. Instead, the interaction between the disk and corona results in the 
evaporation of disk gas, which evacuates the disk. Thus, at low rates of accretion, only hot accretion 
describes the flow, which is similar to the case with a cold gas supply to the disk as described in the 
low/hard state of BHXRBs.
 
\subsection{Spatial distribution of mass flow rate in the disk and corona}

\begin{figure}
\plotone{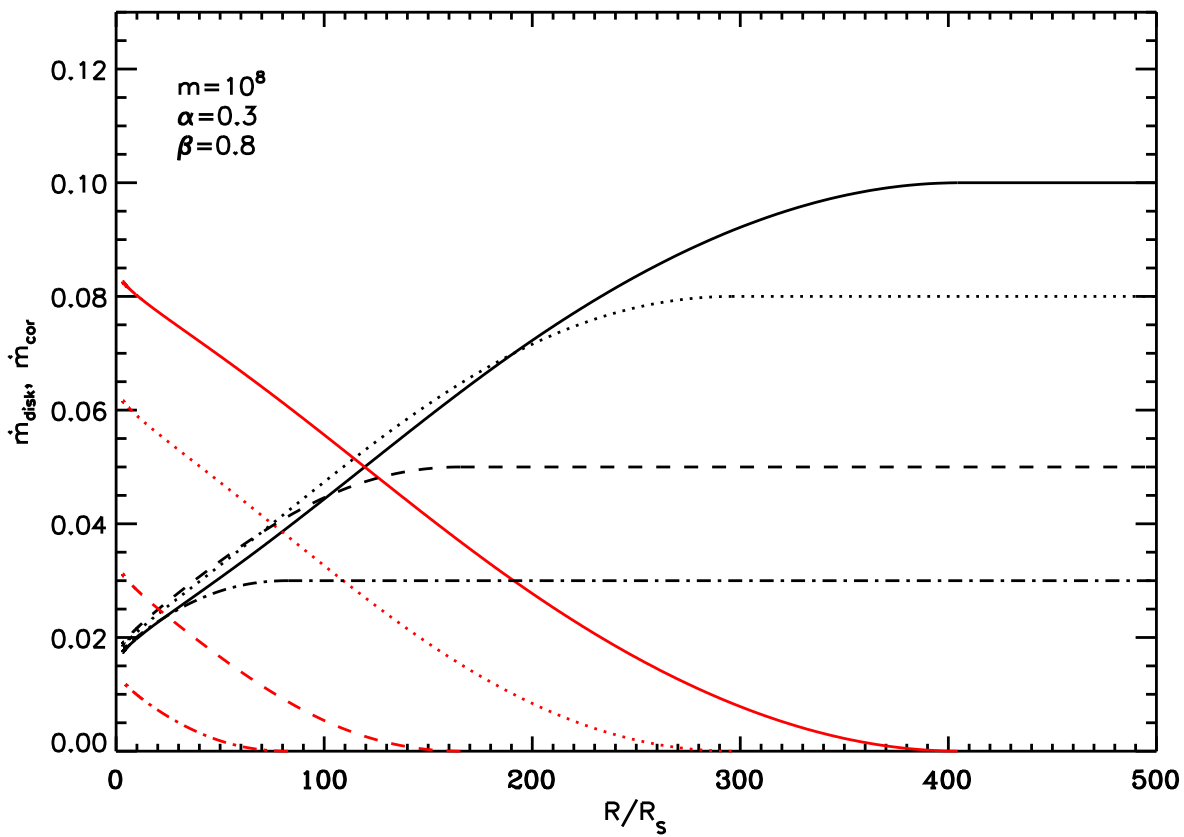}
\caption{\label{f:mdotc-r} The spatial distribution of the mass flow rate in the disk and corona for a black hole 
of $10^8M_\odot$.  Curves in red (black) refer to the mass flow rate in the disk (corona).  From bottom to top, 
the dash-dotted lines refer to a hot mass supply rate of $\dot m=0.03$, dashed lines for $\dot m=0.05$, dotted 
lines for $\dot m=0.08$ and solid lines for $\dot m=0.1$. The mass flow rate in the corona converges to $\dot m\sim 
0.02$ in the innermost region, indicating that a nearly constant corona coexists with a variable disk.}
\end{figure}

The spatial distribution of the mass flow rate in the corona and in the disk is shown in Figure 
\ref{f:mdotc-r} for a black hole of $10^8 \msun$ and hot gas mass flow rates of 0.03, 0.05, 0.08, and 0.1 
(in Eddington units). This mass supply rate is assumed to be captured at large distances, which can be the Bondi radius, and accretes  toward the central black hole.  At a few thousand Schwarzschild radii the temperature of the accreting gas  is rather high compared to the interstellar medium value since  the gravitational potential energy was partially converted into internal energy  via an ADAF within the Bondi radius. In order to locate the critical condensation radius, $R_d$, we start our calculations from 1500 Schwarzschild radii and find that $R_d$ is at distances smaller than $1500R_S$.  In the outer corona ($R>R_d$) evaporation evacuates the presumed disk and eventually the accretion rate is constant. 
With the condensation of gas, the accretion rate 
in the inner corona can be roughly described by a linear function, $\dot m_{\rm cor}\approx 0.02+3\times 10^{-4}{R\over 
R_S}$.  Of importance is the result that the mass flow rate in the innermost region of the corona is nearly 
independent of the mass supply rate, which converges to $\dot m_{\rm cor} \approx 0.02$. Such a characteristic can be 
understood as follows. For a higher mass supply rate, the density in the hot flow is higher, leading to the 
condensation of more gas to the disk by Compton cooling. Thus, the disk emission becomes stronger, which promotes 
more efficient Compton cooling. As a consequence, more hot gas condenses into the disk. This process stabilizes 
as the density of the hot flow decreases as a result of condensation and the electron temperature decreases (by 
efficient Compton cooling) to restrain the further increase of the Compton cooling rate.  Hence, an equilibrium 
is established. We find that the density and mass flow rate remaining in the inner corona is nearly constant for 
the different supply rates.     With hot gas diverted to the disk,  the corona is optically thin even if the mass supply rate is as high as a few tens of percent of the Eddington value.  

\begin{figure}
\plotone{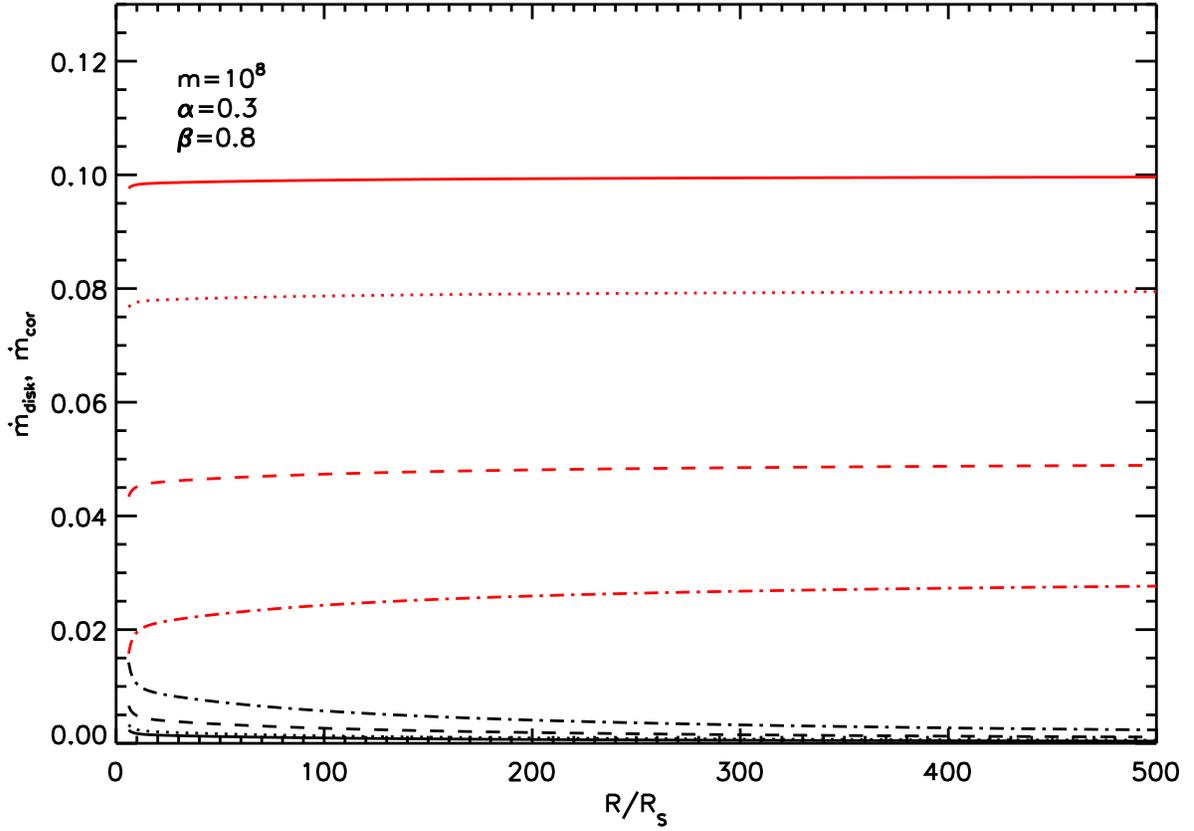}
\caption{\label{f:mdotc2-r} The spatial distribution of the upper limit to the mass flow rate in the corona (black)
and lower limit  in the disk (red) for the case of the mass supply to a thin disk accretion. Symbols are the same 
as in Figure \ref{f:mdotc-r}.}
\end{figure}

Such a feature differs from the results for a cold gas supply to the disk.  As discussed in previous works 
(e.g. Liu et al. 2002b, Meyer-Hofmeister, Liu \& Meyer 2012; Liu et al. 2012), only a small amount of 
cold gas can evaporate into the corona, of which a part condenses back to the disk, resulting in a very weak 
inner coronal flow.  An upper limit to 
the coronal flow rate is estimated in Meyer-Hofmeister et al. (2012) and plotted in Figure \ref{f:mdotc2-r}. It 
reveals that the mass flow rate in the corona is much lower than that in the disk such that very little gas 
remains in the corona, which cannot properly describe the level of X-ray emission observed in AGNs.  

\subsection{Emissions from the disk and the corona}
The gravitational energy released by accretion can be calculated from the standard disk model, that is,
\begin{equation}
L_{acc}=2\int_{R_{\rm in}}^{R_{\rm out}}  {3GM\dot M \left[1-\left(R_{\rm in}\over R\right)^{1/2}\right]\over 8\pi R^3}{2\pi RdR}={1\over 12}\dot M c^2\left[1-{9\over  r_{\rm out}}+2\left(3\over  r_{\rm out}\right)^{3/2}\right],
\end{equation} 
where $\dot M$ is the gas supply rate from a distance $ R_{\rm out}$ with $ r_{\rm out}\equiv R_{\rm out}/R_S$, $R_{\rm in}$ is the ISCO, $R_{\rm in}=3R_S$.  The above equation describes the upper limit to the luminosity emitted from both the disk and the corona,$L_{acc}\approx{1\over 12}\dot M c^2$.  Given a radial distribution of the accretion rate in the corona and disk, 
\begin{equation}
\dot m_{\rm cor}=\left\{ \begin{array}{ll}
ar+b & \mbox{if $r< r_d$}\\
\mbox{$\dot m$} & \mbox{if  $r \ge r_d$}   
      \end{array} \right.
\end{equation}

\begin{equation}
\dot m_{\rm disk}=\left\{ \begin{array}{ll}
\mbox{$\dot m-(ar+b)$} & \mbox{if $r< r_d$}\\
0 & \mbox{if  $r \ge r_d$,}   
      \end{array} \right.
\end{equation}
the luminosity from the disk and corona is, 
\begin{equation}\label{L_disk}
L_{\rm disk}={1\over 12}(\dot m-b)\dot M_{\rm Edd}  c^2\left[1-{9\over  r_{\rm d}}+2\left(3\over  r_{\rm d}\right)^{3/2}\right]-{1\over 12}a\dot M_{\rm Edd}   c^2 \left\{9 ln{r_{\rm d}\over 3}-18\left[{1-\left({3\over r_{\rm d}}\right)^{1/2}}\right]\right\}
\end{equation} 

\begin{equation}\label{L_cor}
L_{\rm cor}={1\over 12}b\dot M_{\rm Edd}  c^2+{1\over 12}a\dot M_{\rm Edd}  c^2 \left\{9 ln{r_{\rm d}\over 3}-18\left[{1-\left({3\over r_{\rm d}}\right)^{1/2}}\right]\right\}+{1\over 12}(\dot m-b)\dot M_{\rm Edd}  c^2 \left[{9\over  r_{\rm d}}-2\left(3\over  r_{\rm d}\right)^{3/2}\right].
\end{equation} 
The first term for the corona luminosity in Eq. \ref{L_cor} is the contribution for a constant rate 
of $b\dot M_{\rm Edd}$ from infinity to the ISCO, whereas the second term corresponds to the 
variable flow with a rate of $ar\dot M_{\rm Edd}$, from $r_d$ to the ISCO. The third term corresponds to 
the contribution in the outer region from infinity to $r_d$ except for the part with a rate of 
$b\dot M_{\rm Edd}$.  

 For a typical mass supply rate, $\dot m= 0.1$, we obtain $a\approx 3\times 10^{-4}$, $b\approx 0.02$, 
and $r_d\approx 400$ from our condensation model.  The ratio of the corona to disk luminosity is $\sim 30\%$ estimated from Eq.(\ref{L_cor}). With a decrease of $\dot m$, condensation takes places at a smaller distance. The ratio of corona and disk 
luminosity increases with decreasing mass supply rate.  We note that the above estimate for the ratio of corona and disk luminosity is a minimum. As the corona density is high, the optical depth of the corona is close to 1, which means more than half of the disk photons are scattered and its energy is contributed to soft and hard X-ray emissions.  On the other hand, the energy released from the corona  partially goes back to the disk. Therefore, we expect the X-ray luminosity 
and disk luminosity to be comparable. 

\begin{figure}
\plotone{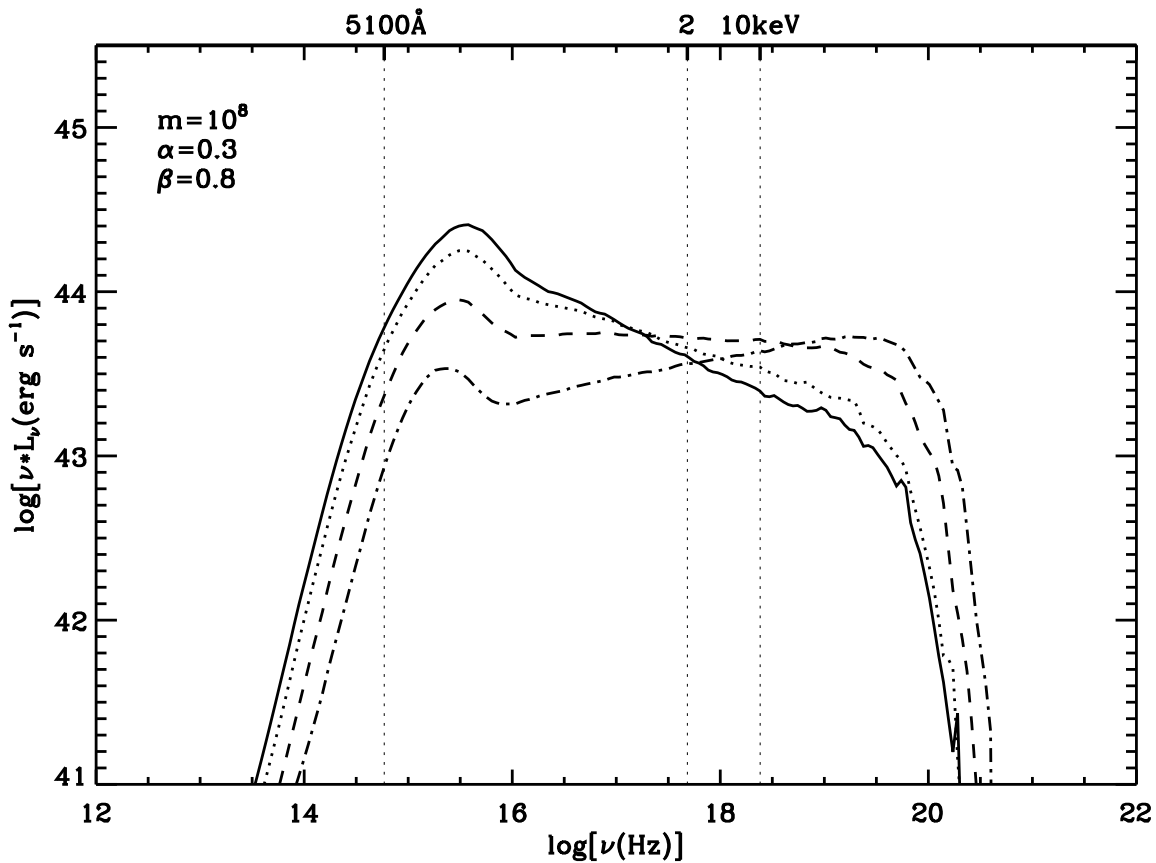}
\caption{\label{f:spectra-mdot} The spectra emitted from a disk and corona with a hot mass supply. Curves from 
bottom to top refer to mass supply rates of 0.03, 0.05, 0.08, and 0.1.  The thin vertical lines denote the 
luminosity at 5100\AA  and 2-10keV, showing the relative strength of the optical to hard X-ray luminosity. 
With an increase of the mass supply rate, the hard X-ray spectrum becomes soft (steep) and the contribution 
of emission from the corona decreases.}
\end{figure}

 The radiation emitted from the disk and corona is calculated numerically for 
$\dot m=0.03,0.05,0.08,0.1$, and the spectra are shown in Figure \ref{f:spectra-mdot}. Here, the spectrum 
is composed of soft photons from the disk, Compton scattering of these soft photons, and the Compton scattering of 
photons produced by the synchrotron and bremsstrahlung processes in the corona.  It can be seen that 
the strength of the emission from the hot corona relative to the disk indeed decreases with an increase of the hot mass 
supply rate or bolometric luminosity. This is a consequence of the increasing condensation rate to the inner disk.  Such a trend is expected for gas supply rates higher than 0.1 if the accretion in the innermost corona remains similar to the case for $\dot m\le 0.1$. The predicted value of $\alpha_{\rm ox}$, which is calculated from the luminosities at $2500\AA$ and 2keV, varies 
from 0.9 to 1.2 with the mass supply rate increasing from 0.03 to 0.1 times the Eddington value respectively.  
The photon index of 
the hard X-ray spectrum increases, varying from 1.9 to 2.3 in the 2-10 keV energy band.  This is a direct result 
of the decrease in the Compton y-parameter associated with the decrease in the electron temperature and slight 
decrease of corona density.

The coronal luminosity, contributed by Bremsstrahlung, synchrotron and inverse Compton scattering, can be comparable with the total disk luminosity even at accretion rate of 0.1. However, the hard X-ray luminosity is obviously smaller than the soft X-ray luminosity, as shown in Figure  \ref{f:spectra-mdot}. 
In order to compare the model prediction with observations, the optical and hard 
X-ray regions are marked in the figure. It can be seen that the 2-10keV X-ray luminosity is of the same order of 
magnitude as the luminosity at $5100\AA$ for accretion rates higher than 0.05. However, the disk radiation peaks 
at frequencies outside this reference band, and the deviations from the luminosity at $5100\AA$ increase 
rapidly with increases of the mass supply rate.  This radiation dominates the whole spectrum at high mass supply 
rates, which leads to an increase of the bolometric correction from the 2-10keV luminosity.

The viscosity and magnetic field also affect the strength of the corona. For example, an increase of the 
viscosity parameter enhances viscous heating, leading to a decrease in gas condensation.  Therefore, more 
gas accretes through the corona, which produces stronger coronal radiation. Similarly, the magnetic field 
contributes additional pressure and hence viscous stress, which also leads to greater heating in the corona 
and less condensation. Hence, both effects lead to a stronger corona.

In summary, the inclusion of an accreting hot/vertically distributed fueling gaseous component leads to the emission 
from the disk and corona in AGN, which significantly differs from the RLO case in BHXRBs. In the latter case, 
the accretion rate in the corona, which is supplied by the disk evaporation, is less than 0.001 for a mass 
supply rate of 0.1 (see Fig. 3), indicating little X-ray emission from the corona whilst the disk emission is 
predominant.

\section{Discussion and Conclusion}

The theoretical description of accretion onto a supermassive black hole in AGN is studied in circumstances 
in which hot gas is present in the central regions of galaxies.   In such an environment matter 
can be gravitationally captured by the supermassive black hole from the interstellar medium or from the mass lost 
by stars.  Here, the hot gas component can arise from the thermalization of the velocity dispersion of stars imprinted 
on stellar mass loss, the mechanical energy injected by this wind loss, by supernovae, or by the effects of 
radiative/mechanical feedback from the AGN. Hot gas flows in, for example, early type galaxies, have been described 
in Pellegrini (2012). In such a case, matter flows inward via a geometrically thick accretion flow at length scales 
corresponding to the Bondi radius, making it possible for hot gas to continuously flow to the central 
black hole without collapsing into a thin disk in luminous AGNs. 

The spatial dependence of this flow in the very innermost regions in accretion disks 
surrounding a supermassive black hole is investigated within the framework of the disk corona interaction 
model.  In order to gain insight of this model, we consider a limiting case where only hot gas is supplied to 
the disk. It is found that the hot gas partially condenses to an underlying cool disk as it flows toward the 
black hole. As a consequence, the mass flow rates of the hot and cool gas components vary with radius.  The 
accretion rates characteristic of the innermost corona are nearly the same for different rates for the hot gas 
supply and correspond to $\dot m\sim 0.02$.  This is in contrast to the conclusion that very little gas 
accretes via a corona as a result of Compton cooling for a cold gas supply at a high accretion rate.  
Therefore, the X-ray emission, which originates from the liberation of gravitational energy of the coronal gas 
and enthalpy of the condensed gas, can be increased relative to a model with only a cold gas supply. 
Hence, our model provides insight in the generation of X-ray radiation and its emission level 
with respect to the optical/UV radiation in HLAGNs without necessarily invoking additional 
heating to the corona in disks fed by a cold gas. The model also predicts that there exists only an 
ADAF flow in LLAGN since the gas evaporates from a thin disk, if existing initially, at low accretion 
rates, quickly evacuating the thin disk.

Calculations of the emission spectrum have been carried out for models applicable to HLAGNs, which reveal 
that the X-ray luminosity can be comparable to the disk luminosity at an accretion rate of $0.1 \dot M_{Edd}$.  
The ratio of the X-ray  to the bolometric luminosity decreases with an increase of the gas supply rate, 
i.e. $\alpha_{\rm ox}$ changes from 0.9 to 1.2 for $\dot m$ from 0.03 to 0.1.
Furthermore, the X-ray photon index increases with increasing mass flow rates, $\Gamma\sim1.9 -2.3$, which is roughly 
 consistent with observations in AGNs. 
  
\subsection{Comparison of the model predictions with gas supply from RLO}

As described in \S 1, previous theoretical studies associated with the disk-corona interaction assume that 
the accreting gas is supplied to a thin disk since the model was primarily developed for the case of BHXRBs 
where gas is transferred from a companion star via RLO.  In this case, a weak corona is maintained by the 
evaporation of matter to the corona.  As matter flows toward the black hole, the evaporation rate increases, 
which diverts matter from the disk to the accreting corona. If the mass transfer rate from the companion star 
is sufficiently low, the accretion disk is all evaporated into the corona at a truncation radius, interior to 
which the accretion flow is replaced by an ADAF.  At intermediate rates, gas evaporated into the corona from 
the outer disk can partially condense back to the disk in the inner region as a consequence of efficient cooling 
through the inverse Compton scattering process.  
For sufficiently high rates, only a small fraction of the disk flow is evaporated into the corona, and the 
thin disk retains its form to the black hole. Hence, the corona is very weak compared to the disk for a very 
high accretion rate.  This implies that the dominant accretion flow varies with the accretion rate, which is in 
good agreement with observational spectra as observed in the low/hard state, the intermediate state, and the 
high/soft state of BHXRBs.

\subsection{The Coronal Outflow?}

 It has been shown (Witt et al. 1997) that a hot accreting corona can produce an outflow.  This was further 
confirmed from detailed vertical structure calculations (Meyer et al. 2000a), where it was found that a 
small fraction of coronal flow can be lost as wind at the upper, transonic boundary. 
Observations of AGN provide support for gas outflow in the neighborhood of a black hole, however, in 
HLAGN as considered in this work, the coronal outflow is expected to be weak. 
This is because the corona in HLAGN is cooler than in traditional ADAFs as a consequence of 
cooling by the inverse Compton process which is unimportant in LLAGN.  This leads to a negative Bernoulli 
parameter and the mass outflow is expected to be significantly reduced relative to the mass inflow rate as compared to 
LLAGN. More importantly, the condensation is so large compared to the outflow that we can neglect outflow in 
determining the coronal features.  Although the mass outflow originating from the corona is not taken into 
account in this work, some outflow from the corona is expected as shown in the work of Witt et al. (1997) and 
Meyer et al. (2000a)).  Outflows driven from the cool disk by radiation pressure in the optically thick 
resonance lines or in failed winds are also expected to be present 
(Czerny \& Hryniewicz 2011, Czerny et al. 2013, Wada 2012).

\subsection{Issues for future study}

In this work,  we explore the flow of a geometrically thick coronal gas component and study its condensation 
to a cool gas component in an optically thick accretion disk surrounding a supermassive black hole. This is a 
limiting case of an AGN.  Given the existence of cool obscuring material and neutral gas in AGNs, a more general 
case where the fuel gas consists of both hot and cold gas should be explored.  In addition,  outflows originating 
from the corona should also be included.  This would be reserved for a future study. 

An estimate of the filling factor of hot gas in the central regions of HLAGNs would be useful 
in determining the viability of the hot flow model in addressing the observational conundrums stemming from 
the application of theoretical accretion disk models for stellar massive black holes in BHXRBs to supermassive 
black holes in AGNs. Such estimates will be observationally challenging, however, given the difficulty in 
separating the X-ray emission of the hot gas fuel from that of the AGN itself. 

\acknowledgments

Financial support for this work is provided by the National Natural Science Foundation of China (grants 
11033007, 11173029, and U1231203) and the Strategic Priority Research Program "The Emergence of Cosmological Structures" of the Chinese Academy of Sciences (Grant No. XDB09000000).  In addition, R.E.T. acknowledges support from the Theoretical Institute for Advanced Research in Astrophysics in the Academia Sinica Institute of Astronomy \& Astrophysics.



\end{document}